\documentclass[twocolumn,showpacs,preprintnumbers]{revtex4}
\usepackage{amsmath}
\usepackage{graphicx}
\usepackage{xcolor}
\usepackage{dcolumn}
\usepackage{bm}

\begin{document}

\title{Topological Spin Density Wave }
\author{Jing He}
\affiliation{Department of Physics, Beijing Normal University, Beijing, 100875 P. R.
China }
\author{Yan-Hua Zong}
\affiliation{Department of Physics, Beijing Normal University, Beijing, 100875 P. R.
China }
\author{Su-Peng Kou}
\thanks{Corresponding author}
\email{spkou@bnu.edu.cn}
\affiliation{Department of Physics, Beijing Normal University, Beijing, 100875 P. R.
China }
\author{Ying Liang}
\affiliation{Department of Physics, Beijing Normal University, Beijing, 100875 P. R.
China }
\author{Shiping Feng}
\affiliation{Department of Physics, Beijing Normal University, Beijing, 100875 P. R.
China }

\begin{abstract}
In this paper, we investigate the topological Hubbard model on honeycomb
lattice. By considering the topological properties of the magnetic state,
new types of quantum states - A-type and B-type topological
spin-density-waves (A-TSDW and B-TSDW) are explored. The low energy physics
is solely determined by its Chern-Simons-Hopf gauge field theories with
different $\mathcal{K}$-matrices. In the formulism of topological field
theory, we found spin-charge separated charge-flux binding effect for A-TSDW
and spin-charge synchronized charge-flux binding effect for B-TSDW. In
addition, we studied the edge states and quantized Hall effect in different
TSDWs.

\end{abstract}

\maketitle

Landau's symmetry breaking paradigm has been very successful as a basis for
understanding the physics of conventional solids including metals and (band)
insulators. In Landau's theory different orders are classified by
symmetries. The phase transitions between one type of ordered phase and
another one (ordered or disordered) are always accompanied by symmetry
breaking. Taking spin-density-wave (SDW) as an example. To describe such
ordered state with spontaneous spin rotation symmetry breaking, one can
define a local order parameter that differs for different SDW states,
(antiferromagnetic (AF) order, ferromagnetic order, ...).

As the first example beyond the Landau's symmetry breaking paradigm, the
integer quantum Hall (IQH) effect is a remarkable achievement in condensed
matter physics\cite{2}. To describe the IQH state, the Chern number or so
called TKNN number, $\mathcal{Q}$, is introduced by integrating over the
Brillouin zone (BZ) of the Berry field strength \cite{thou}. Another example
is the fractional quantum Hall\ (FQH) effect \cite{TSG8259,Laughlinc}, whose
elementary excitations are anyons and the subtle structure is dubbed the
topological order\  \  \cite{Wtop,7}. Recently, a new class of topological
state - topological insulator (TI) is found with the quantized spin Hall
effect\cite{kane,berg}. For all these topological states, elementary
excitations are gapped. Particularly, there is no local order parameter to
characterize them. Instead, the low energy properties of these topological
states can be described by effective Chern-Simons (CS) theories.

In this paper, we will focus on a new type of topological quantum states
with spontaneous spin rotation symmetry breaking - topological SDW (TSDW)
states. Different TSDW states have identical local order parameter - the
staggered magnetization. In order to further classify them, we derive
effective Chern-Simons-Hopf (CSH) theories with different topological
matrices, $\mathcal{K}$-matrices that have been introduced in FQH states\cite%
{Kmat}.

\textit{The topological Hubbard model}: The Hamiltonian of the topological
Hubbard model on honeycomb lattice is given by\cite{he}
\begin{equation}
H=H_{\mathrm{H}}+H^{\prime}+U\sum \limits_{i}\hat{n}_{i\uparrow}\hat {n}%
_{i\downarrow}-\mu \sum \limits_{i,\sigma}\hat{c}_{i\sigma}^{\dagger}\hat {c}%
_{i\sigma}.
\end{equation}
Here $H_{\mathrm{H}}$ is the Hamiltonian of Haldane model\cite{Haldane}
which is given by $H_{\mathrm{H}}=-t\sum \limits_{\left \langle {i,j}%
\right
\rangle ,\sigma}\left( \hat{c}_{i\sigma}^{\dagger}\hat{c}%
_{j\sigma}+h.c.\right) -t^{\prime}\sum \limits_{\left \langle \left \langle {%
i,j}\right
\rangle \right \rangle ,\sigma}e^{i\phi_{ij}}\hat{c}%
_{i\sigma}^{\dagger}\hat {c}_{j\sigma}.$ $t$ and $t^{\prime}$ are the
nearest neighbor and the next nearest neighbor hoppings, respectively. We
introduce a complex phase $\phi_{ij}$ $\left( \left \vert
\phi_{ij}\right
\vert =\frac{\pi}{2}\right) $ to the next nearest neighbor
hopping, of which the positive phase is set to be clockwise. $%
H^{\prime}=\varepsilon \sum \limits_{i\in{A,}\sigma}\hat{c}%
_{i\sigma}^{\dagger}\hat{c}_{i\sigma}-\varepsilon \sum \limits_{i\in{B,}%
\sigma }\hat{c}_{i\sigma}^{\dagger}\hat{c}_{i\sigma}$ denotes an on-site
staggered energy : $\varepsilon$ on $A$ site and $-\varepsilon$ on $B$
site.\ $U$ is the on-site Coulomb repulsion. $\mu$ is the chemical potential
and $\mu=U/2$ at half-filling for our concern in this paper.

When $U$ is zero, the spectrum for free fermions is $E_{\mathbf{k}}=\pm
\sqrt{\left \vert \xi_{k}\right \vert ^{2}+\left(
\xi_{k}^{\prime}+\varepsilon \right) ^{2}}$ where $\left \vert \xi_{\mathbf{k%
}}\right \vert =t\sqrt{3+2\cos{(\sqrt{3}k_{y})}+4\cos{(3k_{x}/2)}\cos{(\sqrt{%
3}k_{y}/2)}}$ and $\xi_{k}^{\prime}=2t^{\prime}\left[ -2\cos \left( {3k_{x}/2%
}\right) \sin({\sqrt{3}k_{y}/2)}+\sin({\sqrt{3}k_{y})}\right] .$ According
to this spectrum, we can see that there exist energy gaps $\Delta_{f1}$, $%
\Delta_{f2}$ near points $\mathbf{k}_{1}=-\frac{2\pi}{3}(1,$ $1/\sqrt{3})$
and $\mathbf{k}_{2}=\frac{2\pi}{3}(1,$ $1/\sqrt{3})$ as $\Delta_{f1}=\left
\vert 2\varepsilon-6\sqrt{3}t^{\prime}\right \vert $ and $%
\Delta_{f2}=2\varepsilon +6\sqrt{3}t^{\prime},$ respectively. In addition,
there exist two phases for this case, the quantum anomalous Hall (QAH) state
and the normal band insulator (NI) state with trivial topological
properties. They are separated by the phase boundary $\Delta_{f1}=0$. In the
QAH state, due to the nonzero TKNN number, $\mathcal{Q}=2$, there exists the
IQH effect with a quantized (charge) Hall conductivity $\sigma_{H}=2e^{2}/h.$

\textit{Mean field (MF) phase diagram and topological quantum phase
transitions}:\textit{\ }Turning on the interaction, the topological Hubbard
model is unstable against AF SDW order that is described by $\langle \hat {c}%
_{i,\sigma}^{\dagger}\hat{c}_{i,\sigma}\rangle=\frac{1}{2}(1+(-1)^{i}\sigma
M)$ where the local order parameter $M$ is the staggered magnetization. We
set $\sigma=+1$ for spin-up and $\sigma=-1$ for spin-down, and then for $%
M\neq0$, the Hamiltonian can be written as $H=H_{\mathrm{H}}+H^{\prime}-\sum
\limits_{i,\sigma}\left( -1\right) ^{i}\Delta_{M}\hat{c}_{i\sigma}^{\dagger
}\sigma_{z}\hat{c}_{i\sigma},$ with $\Delta_{M}=UM/2$. By MF approach, we
obtain the self-consistency equation for $M$ by minimizing the energy at
zero temperature in the reduced BZ as%
\begin{equation}
1=\frac{1}{N_{s}M}\sum \limits_{\mathbf{k}}{[\frac{\xi_{\mathbf{k}}^{\prime
\text{ }}+\Delta_{M}+\varepsilon}{2E_{\mathbf{k}_{1}}}-\frac {\xi_{\mathbf{k}%
}^{\prime \text{ }}-\Delta_{M}+\varepsilon}{2E_{\mathbf{k}_{2}}}{]}}
\end{equation}
where $N_{s}$ is the number of unit cells, $E_{\mathbf{k}_{1}}=\sqrt{(\xi
_{k}^{\prime}+\Delta_{M}+\varepsilon)^{2}+|\xi_{k}|^{2}}$ and $E_{\mathbf{k}%
_{2}}=\sqrt{(\xi_{k}^{\prime}-\Delta_{M}+\varepsilon)^{2}+|\xi_{k}|^{2}}.$

To determine the phase diagram, there are two types of phase transitions :
one is the magnetic phase transition [denoted by $(U/t)_{M}$] between a
magnetic order state with $M\neq0$ and a non-magnetic state with $M=0$, the
other one is the topological phase transition that is characterized by the
condition of zero fermion's energy gaps, $\Delta_{f1}=-6\sqrt{3}%
t^{\prime}+2\varepsilon +2\Delta_{M}=0$ or $\Delta_{f2}=6\sqrt{3}%
t^{\prime}+2\varepsilon-2\Delta _{M}=0$. After determining the phase
boundaries, we plot the phase diagram with five different quantum phases in
FIG.1 for $\varepsilon=0.15$: QAH state, NI, A-type topological SDW state
(A-TSDW), B-type topological SDW state (B-TSDW), and trivial SDW state. In
FIG.2, we also plot the staggered magnetization\ and the energy gaps $%
\Delta_{f1},$ $\Delta_{f2}$\ for the same case.

\begin{figure}[ptbh]
\includegraphics[width = 8.8cm]{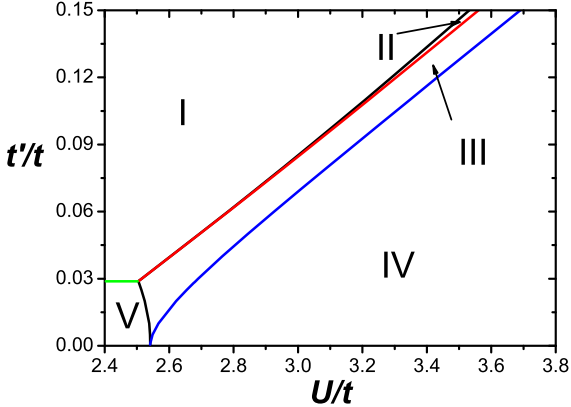}
\caption{(Color online) The phase diagram for the case of $\protect%
\varepsilon=0.15$ : I is QAH state, II is A-TSDW, III is B-TSDW, IV is
trivial SDW, V is NI. The black, red and blue lines are the critical lines
of $(U/t)_{M}$, $(U/t)_{c1}$ and $(U/t)_{c2},$ respectively.}
\end{figure}

\begin{figure}[ptbh]
\includegraphics[width = 10cm]{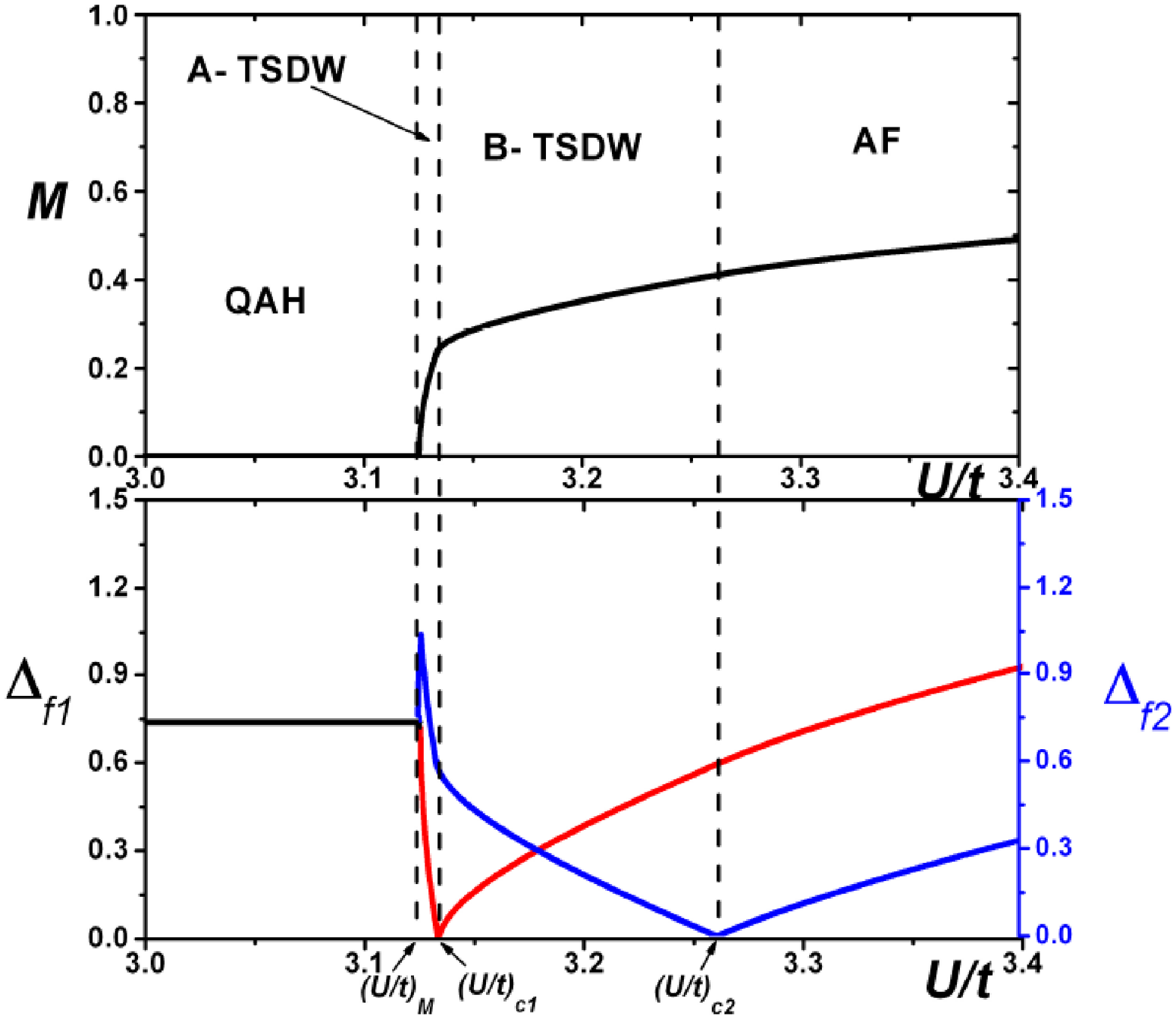}
\caption{(Color online) The staggered magnetization and the energy gaps $%
\Delta_{f1}$, $\Delta_{f2}$ for the fermion excitations at node $\mathbf{k}%
_{1}$, $\mathbf{k}_{2}$ for the case of $\protect \varepsilon=0.15$ and $%
t^{\prime}=0.1$. Below $(U/t)_{M}$, $\Delta _{f1}=\Delta_{f2}$, that is the
black line; above $(U/t)_{M}$, the red line denotes $\Delta_{f1}$ and the
blue line denotes $\Delta_{f2}$.}
\end{figure}

Based on the MF results, the TKNN numbers in A-TSDW, B-TSDW and the trivial
SDW states are $\mathcal{Q}=2$, $\mathcal{Q}=1$, $\mathcal{Q}=0$,
respectively. However, due to the quantum spin fluctuations, the
classification of SDW states by the TKNN number is insufficient to give the
final answer. Instead, a 2-by-2 matrix ($\mathcal{K}$-matrix) plays a key
role in the topological classification of the SDW orders with the same local
order parameter, $M$.

\textit{Induced CSH terms and }$K$\textit{-matrices representation of TSDWs}%
: In this part we will derive the low energy effective theory of (T-)SDW
states by considering quantum fluctuations of effective spin moments based
on a formulation by keeping spin rotation symmetry, $\sigma_{z}\rightarrow
\mathbf{n}\cdot \mathbf{\sigma}$ where $\mathbf{n}$ is the SDW order
parameter, $\left \langle \hat{c}_{i}^{\dagger}\mathbf{\sigma}\hat{c}%
_{i}\right \rangle =M\mathbf{n}$. In this case, the Dirac-like effective
Lagrangian with spin rotation symmetry in the continuum limit can be
obtained as%
\begin{equation}
\mathcal{L}_{f}=\sum_{a}\left[ i\bar{\psi}_{a}\gamma_{\mu}\left(
\partial_{\mu}-iA_{\mu}\right) \psi_{a}+m_{a}\bar{\psi}_{a}\psi_{a}-\delta
\Delta_{M}\bar{\psi}_{a}\mathbf{\sigma}\cdot \mathbf{n}\psi_{a}\right]
\label{f'}
\end{equation}
which describes low energy charged fermionic modes $a=1$ near $\mathbf{k}%
_{1},$ $\bar{\psi}_{1}=\psi_{1}^{\dagger}\gamma_{0}=(%
\begin{array}{llll}
\bar{\psi}_{\uparrow1A}, & \bar{\psi}_{\uparrow1B}, & \bar{\psi}%
_{\downarrow1A}, & \bar{\psi}_{\downarrow1B}%
\end{array}
)$ and $a=2$ near $\mathbf{k}_{2}$, $\bar{\psi}_{2}=\psi_{2}^{\dagger}%
\gamma_{0}=(%
\begin{array}{llll}
\bar{\psi}_{\uparrow2B}, & \bar{\psi}_{\uparrow2A}, & \bar{\psi}%
_{\downarrow2B}, & \bar{\psi}_{\downarrow2A}%
\end{array}
).$ The masses of two-flavor fermions are $m_{1}=\varepsilon-3\sqrt {3}%
t^{\prime}$ and $m_{2}=\varepsilon+3\sqrt{3}t^{\prime}$. $\gamma_{\mu}$ is
defined as $\gamma_{0}=\sigma_{0}\otimes \tau_{z},$ $\gamma_{1}=\sigma
_{0}\otimes \tau_{y},$ $\gamma_{2}=\sigma_{0}\otimes \tau_{x}$ with $\sigma
_{0}=\left(
\begin{array}{ll}
1 & 0 \\
0 & 1%
\end{array}
\right) $. $\tau_{x},$ $\tau_{y},$ $\tau_{z}$ are Pauli matrices. $\delta$ $%
= $ $1$ for $a=1$ and $\delta=-1$ for $a=2$. We have set the Fermi velocity
to be unit $v_{F}=1$.

In\textrm{\ CP}$^{1}$ representation, we may rewrite the effective
Lagrangian of fermions in Eq.(\ref{f'}) as
\begin{equation}
\mathcal{L}_{f}=\sum_{a}\bar{\psi}_{a}^{\prime}\left(
i\gamma_{\mu}\partial_{\mu}+\gamma_{\mu}A_{\mu}-\gamma_{\mu}\sigma_{3}a_{%
\mu}+m_{a}-\delta \Delta_{M}\sigma_{3}\right) \psi_{a}^{\prime}  \label{f2}
\end{equation}
with $\psi_{a}^{^{\prime}}\left( r,\tau \right) =U^{\dagger}\left( r,\tau
\right) \psi_{a}\left( r,\tau \right) $, where $U\left( r,\tau \right) $ is
a local and time-dependent spin \textrm{SU(2)} transformation defined by $%
U^{\dagger}\left( r,\tau \right) \mathbf{n\cdot \sigma}U\left( r,\tau
\right) =\sigma_{3}$. And $a_{\mu}$\ is introduced as an assistant gauge
field as $i\sigma_{3}a_{\mu}\equiv U^{\dagger}\left( r,\tau \right)
\partial_{\mu}U\left( r,\tau \right) .$

An important property of above model in Eq.(\ref{f2}) is the current
anomaly. The vacuum expectation value of the fermionic current $%
J_{a,\sigma}^{\mu }=i\langle \bar{\psi}_{a,\sigma}\gamma^{\mu}\psi_{a,%
\sigma}\rangle$ can be defined by $J_{a,\sigma}^{\mu}=i\{ \gamma^{\mu}[(i%
\hat{D}+im_{a,\sigma }\mathbf{)}^{\dagger}(i\hat{D}+im_{a,\sigma}\mathbf{)}%
]^{-1}(i\hat {D}+im_{a,\sigma}\mathbf{)}^{\dagger}\}$ where $\hat{D}%
=\gamma_{\mu}(\partial_{\mu}-iA_{\mu}+i\sigma a_{\mu})$ and the mass terms
are $m_{a,\sigma}=m_{a}-\delta \Delta_{M}\sigma$. The topological current $%
J_{a,\sigma}^{\mu}$ is obtained to be
\begin{equation}
J_{a,\sigma}^{\mu}=\frac{1}{2}\frac{1}{4\pi}\frac{m_{a,\sigma}}{|m_{a,\sigma
}|}\epsilon^{\mu \nu \lambda}(\partial_{\nu}A_{\lambda}-\sigma \partial_{\nu
}a_{\lambda}).
\end{equation}
Then we derive the CSH terms as $\mathcal{L}_{CSH}=-i\sum_{a,\sigma}(A_{\mu
}-\sigma a_{\mu})J_{a,\sigma}^{\mu}$\cite{redlich,cs}.

To make an explicit description of TSDWs, we introduce the $\mathcal{K}$%
-matrix formulation that has been used to characterize FQH fluids
successfully\cite{Kmat}. Now the CSH term is written as
\begin{equation}
\mathcal{L}_{CSH}=-i\sum_{I,J}\frac{\mathcal{K}_{IJ}}{4\pi }\varepsilon
^{\mu \nu \lambda }a_{\mu }^{I}\partial _{\nu }a_{\lambda }^{J}  \label{cs1}
\end{equation}%
where $\mathcal{K}$ is 2-by-2 matrix, $a_{\mu }^{I=1}=A_{\mu }$ and $a_{\mu
}^{I=2}=a_{\mu }.$ The "charge" of $A_{\mu }$ and $a_{\mu }$ are defined by $%
q$ and $q_{s}$, respectively.

Thus for different SDW orders with the same order parameter $M$, we have
different $\mathcal{K}$-matrices : for A-TSDW order with $m_{1},$ $%
m_{2}>\Delta_{M},$ $\mathcal{K}=\left(
\begin{array}{ll}
2 & 0 \\
0 & 2%
\end{array}
\right) ;$ for B -TSDW order with $m_{2}>\Delta_{M}>m_{1}$, $\mathcal{K}%
=\left(
\begin{array}{ll}
1 & 1 \\
1 & 1%
\end{array}
\right) ;$ for trivial SDW order with $m_{1},$ $m_{2}<\Delta_{M}$, $\mathcal{%
K}=0.$ It is obvious that such topological structure labeled by $\mathcal{K}$%
-matrices is beyond Landau's symmetry breaking paradigm and TKNN number
classification.

In the following parts we will use the following effective model with the
CSH term to learn the topological properties of different SDW orders, $%
\mathcal{L}_{\mathrm{eff}}=\mathcal{L}_{f}+\mathcal{L}_{CSH}$.

\textit{A-TSDW}:\textit{\ }In A-TSDW, an important property is "\emph{%
spin-charge separated} \emph{charge-flux binding}" effect for gauge fields $%
a_{\lambda }$ and $A_{\lambda }$. From the effective CSH Lagrangian in Eq.(%
\ref{cs1}), we get the equations of motion for $a_{\lambda }$ and $%
A_{\lambda },$
\begin{equation}
\frac{1}{\pi }\epsilon ^{\mu \nu \lambda }\partial _{\nu }A_{\lambda
}=-J_{\mu },\text{ }\frac{1}{\pi }\epsilon ^{\mu \nu \lambda }\partial _{\nu
}a_{\lambda }=-J_{s\mu }
\end{equation}%
where $J_{\mu }=i\langle \sum_{a}\bar{\psi}_{a}\gamma _{\mu }\psi
_{a}\rangle $ and $J_{s\mu }=-i\langle \sum_{a}\bar{\psi}_{a}^{\prime
}\gamma _{\mu }\sigma _{3}\psi _{a}^{\prime }\rangle =-i\langle \sum_{a}\bar{%
\psi}_{a}\gamma _{\mu }\mathbf{n\cdot \sigma }\psi _{a}\rangle .$ As a
result, we get the identities $\Phi ^{e}/\pi =-q$ and $\Phi ^{s}/\pi =-q_{s}$%
.

\begin{figure}[ptbh]
\includegraphics[width = 9cm]{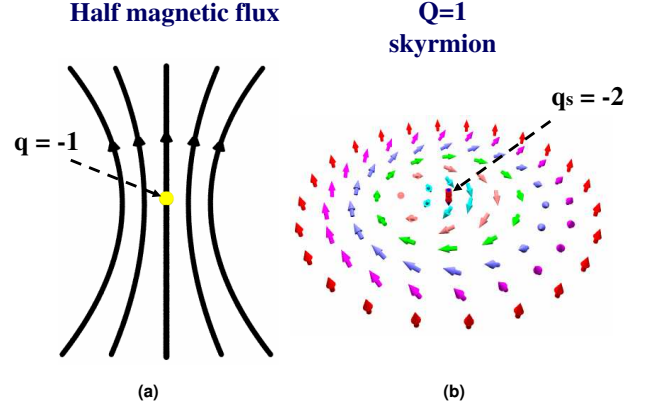}
\caption{(Color online) Spin-charge separated charge-flux binding
effect of A-TSDW : the induced quantum numbers on a half magnetic
flux (a) and those on a $Q=1$ skyrmion (b).}
\end{figure}

On the one hand, the unit magnetic flux of $A_{\lambda}$,
$\Phi^{e}=\int d^{2}\mathbf{r}\epsilon^{0\nu
\lambda}\partial_{\nu}A_{\lambda}$, will bind electric charge
$q=-2$. For example, a half magnetic flux $\Phi^{e}=\pi$ will carry
$q=-1$ electric charge at its core (see FIG.3.(a)). On the other
hand, the unit "magnetic flux" of $a_{\lambda}$ will bind "charge"
$q_{s}=-2$ (see FIG.3.(b)). The "magnetic flux" of $a_{\mu}$ is in
fact the topological spin texture of SDW order (so called $Q$
skyrmion) that is characterized by winding number
$Q=\frac{1}{4\pi}\int \mathbf{d^{2}\mathbf{r}}\epsilon_{0\nu
\lambda }\mathbf{n\cdot \partial_{\nu}\mathbf{n}\times \partial_{\lambda}%
\mathbf{n}}$\cite{pol} from the relation $\partial_{\nu}a_{\lambda}-%
\partial_{\lambda }a_{\nu}=\frac{1}{2}\mathbf{n}\cdot \partial_{\nu}\mathbf{n%
}\times \partial_{\lambda}\mathbf{n}$. Thus in A-TSDW, each skyrmion with $%
\Phi ^{s}=2\pi Q$ "magnetic flux" will accompany by additional "charge"
number $q_{s}=-2Q$. As a result, skyrmion's spin is $S=Q^{2}\Theta/2\pi$
with $\Theta=2\pi.$ For example, by binding $q_{s}=-2$ "charge",\ the spin
of $Q=1$ skyrmion is $S=1$.\

Furthermore, if there exists a tiny easy-plane anisotropic term (for
example, the spin-orbital coupling term), we can define the skyrmion with
fractional winding number - the half skyrmion with $Q=\pm1/2$, of which the
induced "charge" is $q_{s}=\mp1$. However, because the energy of the half
skyrmion in a long range SDW order diverges logarithmically, we cannot treat
it as a real quasiparticle.

In A-TSDW, from the CS term $-\frac{i}{2\pi}\epsilon^{\mu \nu \lambda}A_{\mu
}\partial_{\nu}A_{\lambda}$ and Hopf term $-\frac{i}{2\pi}\epsilon^{\mu \nu
\lambda}a_{\mu}\partial_{\nu}a_{\lambda}$, we find four right-moving
branches of edge excitations instead of two, which are described by the
following one dimension (1D) fermion theory
\begin{equation}
\mathcal{L}_{\text{\textrm{edge}}}=\sum_{\alpha}\psi_{c,\alpha}^{\dag
}(\partial_{t}-v_{c}\partial_{x})\psi_{c,\alpha}+\sum_{\beta}\psi_{s,\beta
}^{\dag}(\partial_{t}-v_{s}\partial_{x})\psi_{s,\beta},
\end{equation}
where $\alpha,\beta=1,2.$ $\psi_{c,\alpha}$ carries a unit of $A_{\mu}$
charge and $\psi_{s,\beta}$ a unit of $a_{\mu}$ charge\cite{7}. That means
we get spin-charge separated edge states : two edge modes only carry
electric charge current; two only carry spin current.

Consequently, we can define two types of Hall conductivities : the quantized
charge Hall conductivity $\sigma_{H}=\lim_{\omega \rightarrow0}{\frac{1}{%
\omega}}\epsilon_{ij}\left \langle J_{i}(\omega,0)J_{j}(-\omega
,0)\right
\rangle $ and the quantized spin Hall conductivity $%
\sigma_{s}=\lim_{\omega \rightarrow0}{\frac{1}{\omega}}\epsilon_{ij}\left
\langle J_{si}(\omega,0)J_{sj}(-\omega,0)\right \rangle .$ Here $J_{i}$
denotes electric current, $J_{i}=i\langle \sum_{a}\bar{\psi}%
_{a}\gamma_{i}\psi _{a}\rangle,$ and $J_{si}$ denotes spin current, $%
J_{si}=-i\langle \sum_{a}\bar{\psi}_{a}\gamma_{i}\mathbf{n\cdot \sigma}%
\psi_{a}\rangle$. From the CSH term, we obtain quantized Hall conductivities
: $\sigma_{H}=\sigma_{s}=2e^{2}/h$ that correspond to the spin-charge
separated edge states\cite{note1}.

\textit{B-TSDW}: In B-TSDW, the equations of motion for $a_{\mu }$ and $%
A_{\mu }$ turn into
\begin{equation}
\frac{1}{2\pi }\epsilon ^{\mu \nu \lambda }\partial _{\nu }A_{\lambda }+%
\frac{1}{2\pi }\epsilon ^{\mu \nu \lambda }\partial _{\nu }a_{\lambda
}=-J_{\mu }
\end{equation}
and
\begin{equation}
\frac{1}{2\pi }\epsilon ^{\mu \nu \lambda }\partial _{\nu }A_{\lambda }+%
\frac{1}{2\pi }\epsilon ^{\mu \nu \lambda }\partial _{\nu }a_{\lambda
}=-J_{s\mu }.
\end{equation}%
As a result, in B-TSDW, due to "\emph{spin-charge synchronized charge-flux
binding}" effect from the identities $\Phi ^{s}/2\pi +\Phi ^{e}/2\pi
=-q=-q_{s}$, the induced electric charge number is always equal to the
induced "charge" number of $a_{\mu }$ on a topological object.

\begin{figure}[ptbh]
\includegraphics[width = 9cm]{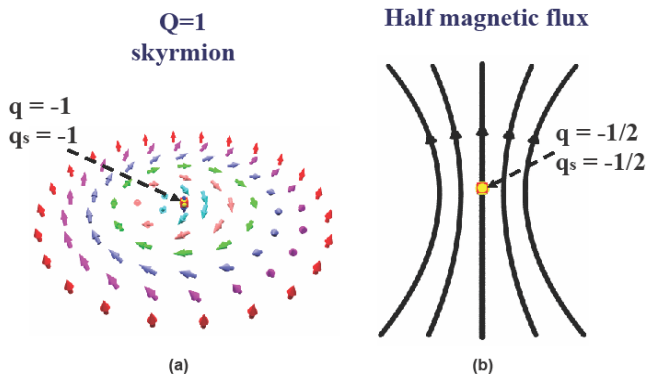}
\caption{(Color online) Spin-charge synchronized charge-flux binding effect
of B-TSDW : the induced quantum numbers on a $Q=1$ skyrmion (a) and those on
a half magnetic flux (b).}
\end{figure}

On the one hand, due to the condition $\Phi^{s}/2\pi=-q=-q_{s},$ $Q$
skyrmion will carry $q=-Q$ electric charge of gauge field $A_{\mu}$ and $%
q_{s}=-Q$ "charge" of gauge field $a_{\mu}$. For example, $Q=1$ skyrmion
carries a unit electric charge $q=-1$ and a unit "charge" $q_{s}=-1$ (see
FIG.4.(a)). With a unit "charge" $q_{s}$, $Q=1$ skyrmion gets half spin and
becomes a charged $S=1/2$ fermion; For a $Q=2$ skyrmion, there exist two
induced charge numbers on it, $q=-2,$ $q_{s}=-2$. Then it becomes a charged $%
S=2$ boson. In addition, for a SDW order with easy-plane anisotropic energy,
we have the half skyrmion with fractional winding number $Q=\pm1/2$, of
which there exist fractional charge numbers, $q=\mp1/2,$ $q_{s}=\mp1/2.$ On
the other hand, using the same approach, from $\Phi^{e}/2\pi=-q=-q_{s},$ we
find that a half quantized magnetic flux $\Phi^{e}=\pi$ will also carry
fractional charge numbers, $q=-1/2,$ $q_{s}=-1/2$ (see FIG.4.(b)).

For B-TSDW, due to a mutual CS term, $-\frac{i}{2\pi}\epsilon^{\mu \nu
\lambda }A_{\mu}\partial_{\nu}a_{\lambda},$ there is no spin-charge
separated edge states. Instead, we have spin-charge synchronized edge
states. Now the charge of $a_{+,\mu}=A_{\mu}+a_{\mu}$ is quantized to an
integer number. Then the effective CSH field theory has one right-moving
edge excitation. The edge excitation is described by the following 1D
fermion theory $\mathcal{L}_{\text{\textrm{edge}}}=\tilde{\psi}%
^{\dag}(\partial_{t}-\tilde{v}\partial _{x})\tilde{\psi}$ where $\tilde{\psi}
$ carries a unit of $a_{+,\mu}$ "charge"\cite{7}. Then we may define the
spin-charge synchronized Hall conductivity as $\tilde{\sigma}=\lim_{\omega
\rightarrow0}{\frac{1}{\omega}}\epsilon_{ij}\left \langle \tilde{J}%
_{i}(\omega,0)\tilde{J}_{j}(-\omega ,0)\right \rangle $ with $\tilde{J}%
_{i}=i\langle \sum_{a}\bar{\psi }_{a,\downarrow}^{\prime}\gamma_{i}\psi_{a,%
\downarrow}^{\prime}\rangle =i\langle \sum_{a}\bar{\psi}_{a}\gamma_{i}(1-%
\mathbf{n\cdot \sigma)}\psi _{a}/2\rangle$ and get a quantized spin-charge
synchronized Hall conductivity : $\tilde{\sigma}=\frac{e^{2}}{h}$ that
corresponds to the edge state. That means an electric field can drive both a
quantized electric charge current and a quantized spin current. In
particular, such a quantized spin-charge synchronized Hall effect is \emph{%
not} QAH effect for electrons with confined spin-charge degrees of freedom
in QAH state.

\textit{Conclusion}: In this paper, we investigate the topological Hubbard
model on honeycomb lattice. New types of quantum states - A-TSDW and B-TSDW
states are explored which bear an identical staggered magnetization $M$ as
the local order parameter. To characterize different TSDWs, we introduce $%
\mathcal{K}$-matrices to denote different CSH gauge theories. In the
formulism of the CSH gauge field theories with different $\mathcal{K}$%
-matrices, we found spin-charge separated charge-flux binding effect in
A-TSDW and spin-charge synchronized charge-flux binding effect in B-TSDW. In
addition we studies the edge states and corresponding quantized Hall effect
in TSDWs. Our findings suggest that although there exists spontaneous spin
rotation symmetry breaking, the TSDWs are beyond Landau's paradigm.

\begin{acknowledgments}
This word is supported by SRFDP, NFSC Grant No. 10874017 and 10774015,
National Basic Research Program of China (973 Program) under the grant No.
2011CB921803, 2011cba00102.
\end{acknowledgments}

\subsection{The detailed calculations of the Chern-Simons-Hopf (CSH) terms
in Eq.(6)}

Firstly we calculate the induced Chern-Simons (CS) term of a one flavor
fermionic-$\sigma$ model. The Lagrangian of one flavor fermionic-$\sigma$
model is written as
\begin{equation*}
\mathcal{L}=i\bar{\psi}\gamma_{\mu}(\partial_{\mu}-ib_{\mu})\psi+m\bar{\psi }%
\psi
\end{equation*}
where $m$ is a fermion mass. To obtain the induced CS term, we integrating
over fermions and get
\begin{equation*}
S_{\mathrm{eff}}=\ln \mathcal{Z}
\end{equation*}
where
\begin{equation}
\mathcal{Z}=\int[d\psi][d\bar{\psi}]\exp{(-\int dx\mathcal{L}})\mathbf{.}
\notag
\end{equation}
The one fermion loop effective action becomes
\begin{align}
S_{\mathrm{eff}} & =\ln \det(i\gamma_{\mu}\partial_{\mu}+\gamma_{\mu}b_{\mu
}+m) \\
& =\, \mathrm{tr}\log(i\gamma_{\mu}\partial_{\mu}+m)+\mathrm{tr}({\frac {1}{%
i\gamma_{\mu}\partial_{\mu}+m}}\gamma_{\mu}b_{\mu})  \notag \\
& +\frac{1}{2}\, \mathrm{tr(}{\frac{1}{i\gamma_{\mu}\partial_{\mu}+m}}\cdot
\gamma_{\mu}b_{\mu}{\frac{1}{i\gamma_{\mu}\partial_{\mu}+m}}\gamma_{\mu
}b_{\mu})+\dots  \notag
\end{align}
Then the quadratic term of $b_{\mu}$ in the effective action is
\begin{equation}
S_{\mathrm{eff}}=\frac{1}{2}\int{\frac{d^{3}p}{(2\pi)^{3}}}\left[ b^{\mu
}(-p)D^{\mu \nu}b^{\nu}(p)\right]  \label{quadr}
\end{equation}
where $D^{\mu \nu}$ is
\begin{equation}
D^{\mu \nu}=\int{\frac{d^{3}k}{(2\pi)^{3}}}\, \mathrm{tr}\left[ \gamma^{\mu
}\,{\frac{p_{\mu}\gamma_{\mu}+k_{\mu}\gamma_{\mu}-m}{(p+k)^{2}+m^{2}}}\,
\gamma^{\nu}\,{\frac{k_{\mu}\gamma_{\mu}-m}{k^{2}+m^{2}}}\right] .
\end{equation}
Under $\mathrm{tr}(\gamma^{\mu}\gamma^{\nu}\gamma^{\lambda})=-2\epsilon
^{\mu \nu \lambda},$ we obtain
\begin{align}
D^{\mu \nu}(p,m) & =\epsilon^{\mu \nu \lambda}p_{\lambda}2m\int{\frac{d^{3}k%
}{(2\pi)^{3}}}{\frac{1}{[(p+k)^{2}+m^{2}][k^{2}+m^{2}]}} \\
& =\epsilon^{\mu \nu \lambda}p_{\lambda}\frac{1}{2\pi}\frac{m}{|p|}\arcsin({%
\frac{|p|}{\sqrt{p^{2}+4m^{2}}}}).  \notag
\end{align}
In the long wavelength limit, ($\frac{p}{m}\rightarrow0$), due to $\Theta
\sim \, \frac{1}{4\pi}\frac{m}{|m|}$, we get an induced CS term as
\begin{equation}
\mathcal{L}_{\mathrm{eff}}=-\frac{i}{8\pi}\frac{m}{|m|}\epsilon^{\mu \nu
\lambda}b_{\mu}\partial_{\nu}b_{\lambda}
\end{equation}

Next we calculate the induced CSH term in Eq.(6) in the paper that denotes a
four-component fermionic model
\begin{equation}
\mathcal{L}_{1\uparrow}=\bar{\psi}_{1\uparrow}^{\prime}\left( i\gamma^{\mu
}\partial_{\mu}+\gamma^{\mu}A_{\mu}-\gamma^{\mu}a_{\mu}+m_{1}-\Delta
_{M}\right) \psi_{1\uparrow}^{\prime}.  \notag
\end{equation}%
\begin{equation}
\mathcal{L}_{1\downarrow}=\bar{\psi}_{1\downarrow}^{\prime}\left(
i\gamma^{\mu}\partial_{\mu}+\gamma^{\mu}A_{\mu}+\gamma^{\mu}a_{\mu}+m_{1}+%
\Delta_{M}\right) \psi_{1\downarrow}^{\prime}.  \notag
\end{equation}%
\begin{equation}
\mathcal{L}_{2\uparrow}=\bar{\psi}_{2\uparrow}^{\prime}\left( i\gamma^{\mu
}\partial_{\mu}+\gamma^{\mu}A_{\mu}-\gamma^{\mu}a_{\mu}+m_{2}+\Delta
_{M}\right) \psi_{2\uparrow}^{\prime}.  \notag
\end{equation}%
\begin{equation}
\mathcal{L}_{2\downarrow}=\bar{\psi}_{2\downarrow}^{\prime}\left(
i\gamma^{\mu}\partial_{\mu}+\gamma^{\mu}A_{\mu}+\gamma^{\mu}a_{\mu}+m_{2}-%
\Delta_{M}\right) \psi_{2\downarrow}^{\prime}.  \notag
\end{equation}
Integrating $\psi_{1\uparrow}^{\prime},$ we get the induced CS term
\begin{equation*}
\mathcal{L}_{1\uparrow}(A_{\mu})=-\frac{i}{8\pi}\frac{m_{1}-\Delta_{M}}{%
|m_{1}-\Delta_{M}|}\epsilon^{\mu \nu \lambda}(A_{\mu}-a_{\mu})\partial_{\nu
}(A_{\lambda}-a_{\lambda});
\end{equation*}
Integrating $\psi_{1\downarrow}^{\prime},$ we get the induced CS term
\begin{equation*}
\mathcal{L}_{1\downarrow}(A_{\mu})=-\frac{i}{8\pi}\frac{m_{1}+\Delta_{M}}{%
|m_{1}+\Delta_{M}|}\epsilon^{\mu \nu \lambda}(A_{\mu}+a_{\mu})\partial_{\nu
}(A_{\lambda}+a_{\lambda});
\end{equation*}
Integrating $\psi_{2\uparrow}^{\prime},$ we get the induced CS term
\begin{equation*}
\mathcal{L}_{2\uparrow}(A_{\mu})=-\frac{i}{8\pi}\frac{m_{2}+\Delta_{M}}{%
|m_{2}+\Delta_{M}|}\epsilon^{\mu \nu \lambda}(A_{\mu}-a_{\mu})\partial_{\nu
}(A_{\lambda}-a_{\lambda});
\end{equation*}
Integrating $\psi_{2\downarrow}^{\prime},$ we get the induced CS term
\begin{equation*}
\mathcal{L}_{2\downarrow}(A_{\mu})=-\frac{i}{8\pi}\frac{m_{2}-\Delta_{M}}{%
|m_{2}-\Delta_{M}|}\epsilon^{\mu \nu \lambda}(A_{\mu}+a_{\mu})\partial_{\nu
}(A_{\lambda}+a_{\lambda}).
\end{equation*}

Then we get the total induced CSH term as
\begin{equation*}
\mathcal{L}(A_{\mu})=\mathcal{L}_{1\uparrow}(A_{\mu})+\mathcal{L}%
_{1\downarrow}(A_{\mu})+\mathcal{L}_{2\uparrow}(A_{\mu})+\mathcal{L}%
_{2\downarrow}(A_{\mu}).
\end{equation*}
(i) For the case of $m_{1},$ $m_{2}>\Delta_{M},$ we get
\begin{align*}
\mathcal{L}(A_{\mu}) & =\mathcal{L}_{1\uparrow}(A_{\mu})+\mathcal{L}%
_{1\downarrow}(A_{\mu})+\mathcal{L}_{2\uparrow}(A_{\mu})+\mathcal{L}%
_{2\downarrow}(A_{\mu}) \\
& =-\frac{i}{8\pi}\epsilon^{\mu \nu \lambda}(A_{\mu}+a_{\mu})\partial_{\nu
}(A_{\lambda}+a_{\lambda}) \\
& -\frac{i}{8\pi}\epsilon^{\mu \nu \lambda}(A_{\mu}-a_{\mu})\partial_{\nu
}(A_{\lambda}-a_{\lambda}) \\
& -\frac{i}{8\pi}\epsilon^{\mu \nu \lambda}(A_{\mu}-a_{\mu})\partial_{\nu
}(A_{\lambda}-a_{\lambda}) \\
& -\frac{i}{8\pi}\epsilon^{\mu \nu \lambda}(A_{\mu}+a_{\mu})\partial_{\nu
}(A_{\lambda}+a_{\lambda}) \\
& =\frac{-i}{2\pi}\epsilon^{\mu \nu \lambda}A_{\mu}\partial_{\nu}A_{\lambda
}+\frac{-i}{2\pi}\epsilon^{\mu \nu \lambda}a_{\mu}\partial_{\nu}a_{\lambda};
\end{align*}
(ii) For the case of $m_{2}>\Delta_{M}>m_{1},$ we get
\begin{align*}
\mathcal{L}(A_{\mu}) & =\mathcal{L}_{1\uparrow}(A_{\mu})+\mathcal{L}%
_{1\downarrow}(A_{\mu})+\mathcal{L}_{2\uparrow}(A_{\mu})+\mathcal{L}%
_{2\downarrow}(A_{\mu}) \\
& =-\frac{i}{8\pi}\epsilon^{\mu \nu \lambda}(A_{\mu}+a_{\mu})\partial_{\nu
}(A_{\lambda}+a_{\lambda}) \\
& +\frac{i}{8\pi}\epsilon^{\mu \nu \lambda}(A_{\mu}-a_{\mu})\partial_{\nu
}(A_{\lambda}-a_{\lambda}) \\
& -\frac{i}{8\pi}\epsilon^{\mu \nu \lambda}(A_{\mu}-a_{\mu})\partial_{\nu
}(A_{\lambda}-a_{\lambda}) \\
& -\frac{i}{8\pi}\epsilon^{\mu \nu \lambda}(A_{\mu}+a_{\mu})\partial_{\nu
}(A_{\lambda}+a_{\lambda}) \\
& =\frac{-i}{4\pi}\epsilon^{\mu \nu \lambda}A_{\mu}\partial_{\nu}A_{\lambda
}+\frac{-i}{2\pi}\epsilon^{\mu \nu \lambda}A_{\mu}\partial_{\nu}a_{\lambda }+%
\frac{-i}{4\pi}\epsilon^{\mu \nu \lambda}a_{\mu}\partial_{\nu}a_{\lambda};
\end{align*}
(iii) For the case of $m_{1},$ $m_{2}<\Delta_{M},$ we get
\begin{align*}
\mathcal{L}(A_{\mu}) & =\mathcal{L}_{1\uparrow}(A_{\mu})+\mathcal{L}%
_{1\downarrow}(A_{\mu})+\mathcal{L}_{2\uparrow}(A_{\mu})+\mathcal{L}%
_{2\downarrow}(A_{\mu}) \\
& =-\frac{i}{8\pi}\epsilon^{\mu \nu \lambda}(A_{\mu}+a_{\mu})\partial_{\nu
}(A_{\lambda}+a_{\lambda}) \\
& +\frac{i}{8\pi}\epsilon^{\mu \nu \lambda}(A_{\mu}-a_{\mu})\partial_{\nu
}(A_{\lambda}-a_{\lambda}) \\
& -\frac{i}{8\pi}\epsilon^{\mu \nu \lambda}(A_{\mu}-a_{\mu})\partial_{\nu
}(A_{\lambda}-a_{\lambda}) \\
& +\frac{i}{8\pi}\epsilon^{\mu \nu \lambda}(A_{\mu}+a_{\mu})\partial_{\nu
}(A_{\lambda}+a_{\lambda}) \\
& =0.
\end{align*}

Finally, by the $\mathcal{K}$-matrix, the CSH term is written as
\begin{equation}
\mathcal{L}_{CSH}=-i\sum_{I,J}\frac{\mathcal{K}_{IJ}}{4\pi}\varepsilon^{\mu
\nu \lambda}a_{\mu}^{I}\partial_{\nu}a_{\lambda}^{J}  \label{cs}
\end{equation}
where $\mathcal{K}$ is 2-by-2 matrix, $a_{\mu}^{I=1}=A_{\mu}$ and $a_{\mu
}^{I=2}=a_{\mu}.$ Thus for different SDW orders with the same order
parameter $M$, we have different $\mathcal{K}$-matrices : for $m_{1},$ $%
m_{2}>\Delta _{M},$ $\mathcal{K}=\left(
\begin{array}{ll}
2 & 0 \\
0 & 2%
\end{array}
\right) ;$ for $m_{2}>\Delta_{M}>m_{1}$, $\mathcal{K}=\left(
\begin{array}{ll}
1 & 1 \\
1 & 1%
\end{array}
\right) ;$ for $m_{1},$ $m_{2}<\Delta_{M}$, $\mathcal{K}=0.$


\begin{thebibliography}{99}
\bibitem{note1} To define a quantized spin Hall conductivity, we need to add
an additional easy-axis anisotropic energy term to open a small spin gap.
\end{thebibliography}
\end{document}